\documentclass[usenatbib,usegraphicx]{mn2e}
\usepackage{amsfonts}
\usepackage{amsmath}
\usepackage{amssymb}
\usepackage{comment}
\usepackage{color}

\title[Major-merger-driven size growth]
    {The role of major mergers in the size growth of {\color{black}intermediate-mass} spheroids\vspace{-0.2in}}

\author[Sugata Kaviraj et al.]
{S. Kaviraj,$^{1,2}$ M. Huertas-Company,$^{3,4}$ S. Cohen,$^{5}$
S. Peirani,$^{6}$ R. A. Windhorst$^{5}$, \newauthor R. W.
O'Connell,$^{7}$ J. Silk,$^{6,2}$ M. A. Dopita,$^{8,9}$ N. P.
Hathi,$^{10}$ A. M. Koekemoer,$^{11}$
\newauthor S. Mei,$^{3,4}$ M. Rutkowski,$^{5}$ R. E. Ryan,$^{11}$ and F. Shankar$^{12,3}$\vspace{0.05in}\\
$^{1}$Centre for Astrophysics Research, University of Hertfordshire, College Lane, Hatfield, Herts, AL10 9AB, UK\\
$^{2}$Department of Physics, University of Oxford, Keble Road, Oxford, OX1 3RH, UK\\
$^{3}$GEPI, Paris Observatory, 77 Avenue Denfert Rochereau, F-75014 Paris, France\\
$^{4}$University Denis Diderot, 4 Rue Thomas Mann, F-75205 Paris, France\\
$^{5}$School of Earth and Space Exploration, Arizona State University, Tempe, AZ 85287-1404, USA\\
$^{6}$Institut d'Astrophysique de Paris, 98 bis boulevard Arago, 75014 Paris, France\\
$^{7}$Department of Astronomy, University of Virginia, Charlottesville, VA 22904-4325, USA\\
$^{8}$Research School of Physics and Astronomy, The Australian National University, ACT 2611, Australia\\
$^{9}$King Abdulaziz University, Astronomy Department, Faculty of Science, Jeddah, Saudi Arabia\\
$^{10}$Carnegie Observatories, 813 Santa Barbara Street, Pasadena, California, 91101, USA\\
$^{11}$Space Telescope Science Institute, 3700 San Martin Drive,
Baltimore, MD 21218, USA\\
$^{12}$School of Physics \& Astronomy, University of Southampton,
Highfield, Southampton SO17 1BJ, UK}

\begin{document}

\maketitle

\def \aj {AJ}
\def \mnras {MNRAS}
\def \pasp {PASP}
\def \apj {ApJ}
\def \apjs {ApJS}
\def \apjl {ApJL}
\def \aap {A\&A}
\def \nat {Nature}
\def \araa {ARAA}
\def \iaucirc {IAUC}
\def \aaps {A\&A Suppl.}
\def \qjras {QJRAS}
\def \na {New Astronomy}
\def \aapr {A\&ARv}
\def\lesssim{\mathrel{\hbox{\rlap{\hbox{\lower4pt\hbox{$\sim$}}}\hbox{$<$}}}}
\def\gtrsim{\mathrel{\hbox{\rlap{\hbox{\lower4pt\hbox{$\sim$}}}\hbox{$>$}}}}


\begin{abstract}
We study of the role of `major' mergers (mass ratios $>1:4$) in
driving size growth in high-redshift ($1<z<2$) spheroidal galaxies
(SGs) with stellar masses between $10^{9.5}$ M$_{\odot}$ and
$10^{10.7}$ M$_{\odot}$. This is a largely unexplored mass range
at this epoch, containing the progenitors of more massive SGs on
which the bulk of the size-evolution literature is based. We
visually split our SGs into systems that are relaxed and those
that exhibit tidal features indicative of a recent merger.
Numerical simulations indicate that, given the depth of our
images, only tidal features due to major mergers will be
detectable at these epochs (features from minor mergers being too
faint), making the disturbed SGs a useful route for estimating
major-merger-driven size growth. The disturbed SGs are offset in
size from their relaxed counterparts, lying close to the upper
envelope of the local size -- mass relation. The mean size ratio
of the disturbed SGs to their relaxed counterparts is $\sim$2.
Combining this observed size growth with empirical major-merger
histories from the literature suggests that {\color{black} the
size evolution of a significant fraction (around two-thirds) of
SGs \emph{in this mass range} could be driven by major mergers.
If, as is likely, our galaxies are progenitors of more massive
($M_*>10^{10.7}$ M$_{\odot}$) SGs at $z<1$, then major mergers are
also likely to play an important role in the size growth of at
least some massive SGs in this mass range.}
\end{abstract}




\section{Introduction}
Massive spheroidal galaxies (SGs) dominate the local stellar mass
density \citep[e.g.][]{Kaviraj2014a}, making them unique probes of
galaxy evolution over cosmic time. Consensus has recently moved
away from the classical notion that SGs are old and
passively-evolving systems. While spectro-photometric studies have
revealed widespread recent star formation and a surprising
diversity of formation epochs
\citep[e.g.][]{Kaviraj2007a,Kaviraj2008,Trager2008}, structural
studies have indicated strong size evolution, with SGs at $z
\sim3$ being 3--5 times smaller than their counterparts today
\citep[e.g.][]{Daddi2005,Trujillo2006,Buitrago2008,Cimatti2012,Ryan2012,Huertas-Company2013}.

While SG sizes do increase over cosmic time, the processes that
drive this evolution remain debated. Various theoretical scenarios
have been proposed to explain the size growth, including dry major
and minor merging \citep[e.g.][]{Bournaud2007,Oser2012} and
secular mechanisms, such as adiabatic expansion driven by stellar
mass loss or strong AGN feedback \citep[e.g.][but see e.g.
Trujillo et al. 2011]{Fan2008,Damjanov2009}, {\color{black}with
the bulk of the literature focussing on the evolution of massive
(M$_{\odot} > 10^{10.7}$ M$_{\odot}$) spheroids. Recent work
indicates that, in this mass range (and particularly in the
redshift range $z<1$), the combination of major and minor mergers
may account for much of the size growth
\citep[e.g.][]{Bluck2012,Newman2012,Lopez2012}, although some of
the observed size evolution is driven by the appearance of large
newly-quenched galaxies - the so-called `progenitor bias'
\citep[e.g.][]{VanderWel2009,Carollo2013}.}

While theoretical explanations based on merging have been proposed
to explain SG size evolution, it is clearly desirable to have an
\emph{empirical} estimate of size growth that can be induced by
merging, at the epoch where much of the growth is expected to take
place ($z>1$). {\color{black}This is particularly useful for the
mass range M$_{\odot} < 10^{10.7}$ M$_{\odot}$ which is not
typically the focus of the bulk of the literature.} Such an
empirical estimate is both a useful observational result and a
quantitative constraint on merger-driven scenarios that aim to
explain the size evolution of the SG population {\color{black}in
this mass range.}

\begin{figure}
\begin{center}
\includegraphics[width=3in]{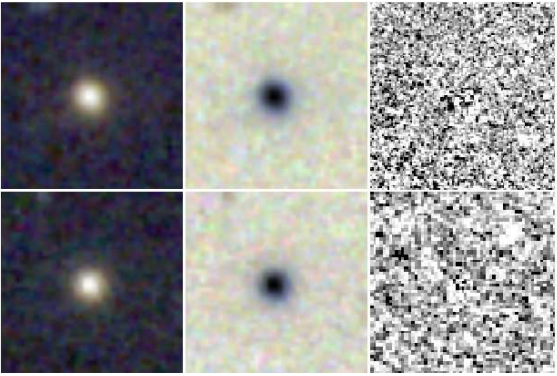}
\caption{Examples of relaxed SGs in our dataset. We show both the
$YJH$ colour composite image (left column) and its negative
(middle column). The right-hand column indicates the residual
image from \texttt{GALFIT}.} \label{fig:relaxed_SG_examples}
\end{center}
\end{figure}

The identification of SGs and measurement of their sizes at $z>1$
greatly benefits from \emph{Hubble Space Telescope (HST)} imaging
in the near-infrared (NIR). At these epochs, the NIR corresponds
to the rest-frame optical, which traces the underlying stellar
population of the galaxy and not just the UV-emitting star-forming
regions. New NIR surveys using HST's Wide Field Camera 3 (WFC3),
such as the WFC3 Early-Release Science (ERS) programme
\citep{Windhorst2011}, are providing unprecedented large-scale
access to NIR data at $z>1$, making them ideal datasets for such a
study. In this Letter, we use 80 SGs at $1<z<2$, drawn from the
ERS, to calculate an empirical estimate of size growth due to
mergers with mass ratios $>$1:4 (`major' mergers) and explore the
proportion of SG size evolution that may be attributable to the
major-merger process in the stellar mass range $10^{9.5}$
M$_{\odot}$$<M_*<10^{10.7}$ M$_{\odot}$. This is an unexplored
mass range at this epoch, containing the progenitors of more
massive galaxies, on which the SG size-evolution literature is
largely based.

\citet[][see their \S4]{Kaviraj2013a} have recently used
hydrodynamical simulations to show that the presence of tidal
features around ERS galaxies in the redshift range $1<z<2$
indicates a recent major merger (mass ratios $<$1:4). The ERS
images are too shallow to reveal the fainter tidal features
produced by minor mergers. Thus, separating the ERS SGs into those
that are relaxed and those that are tidally disturbed provides a
route to estimating the structural impact of major mergers on
these systems. While we cannot track the evolution of individual
galaxies, we can use the relaxed and disturbed SG populations as a
whole to derive a mean statistical estimate of the size growth
induced by the major-merger process. Combining this with the
typical major-merger histories of massive galaxies then enables us
to explore the potential contribution of this process to SG size
evolution over cosmic time.

This Letter is organised as follows. In \S2, we describe the ERS galaxy
sample used in this study, the selection of relaxed and disturbed SGs via
visual inspection and the derivation of galaxy stellar masses, rest-frame
photometry and sizes that underpin our analysis. In \S3, we explore the
effective radii of galaxies in these two SG subpopulations and discuss the
role of major mergers in driving size growth. We summarise our results in
\S4. Throughout, we employ the WMAP7 cosmological parameters
\citep{Komatsu2011} and photometry in the AB system \citep{Oke1983}.


\section{Data}

\subsection{The WFC3 Early-Release Science programme}
The WFC3 ERS programme has imaged around one-third of the
GOODS-South field \citep{Giavalisco2004} using the UVIS and IR
channels of the WFC3, with a total exposure time of 104 orbits.
The observations, data reduction, and instrument performance are
described in detail in Windhorst et al. (2011) and summarised
here. The UVIS data (40 orbits) covers $\sim$55 arcmin$^2$, in
each of the F225W, F275W and F336W filters, with relative exposure
times of 2:2:1. The NIR data (60 orbits) covers $\sim$45
arcmin$^2$ using the F098M ($Y_s$), F125W ($J$), and F160W ($H$)
filters, with equal exposure times of 2 orbits per filter.
Together, the data provide 10-band HST panchromatic coverage over
0.2 - 1.7 $\mu$m, with 5$\sigma$ point-source depths of $AB \sim
26.4$ mag and $AB \sim27.5$ mag in the UV and NIR respectively.

In this paper we study 80 SGs in the ERS that are brighter than
$H(AB)=24$ and have either spectroscopic or photometric redshifts
in the range $1<z<2$. Photometric redshifts are calculated by
applying the EAZY code \citep{Brammer2008} on the 10-band WFC3/ACS
photometric catalogue. Spectroscopic redshifts are drawn from the
literature, from spectra taken using the Very Large Telescope
\citep{LeFevre2004,Szokoly2004,Popesso2009}, the Keck telescopes
\citep{Strolger2004} and the HST ACS grism
\citep{Daddi2005,Pasquali2006}. For the analysis that follows,
spectroscopic redshifts (available for 14\% of our galaxies) are
always used where available.


\subsection{Selection of spheroids via visual inspection}
Following \citet{Kaviraj2013a}, SGs are selected via visual
inspection of $YJH$ composite images, scaled using the asinh
method of \citet{Lupton2004}. Using multi-filter composites --
instead of monochrome images -- maximises the rest-frame optical
information in the image, facilitating the identification of tidal
features. We restrict our study to galaxies brighter than
$H(AB)=24$. Past work that has used visual inspection of HST
images for morphological classification has typically employed
rest-frame optical imaging with similar or fainter
surface-brightness limits compared to the images employed here
\citep[e.g.][]{Robaina2009,Kaviraj2011}. Where appropriate, we
flag the presence of tidal features, thus splitting our sample
into `relaxed' SGs (R-SGs) and `disturbed' SGs (D-SGs). As
mentioned in the introduction, the D-SGs are likely to have
experienced recent major mergers, since the ERS images are not
deep enough to reveal the fainter tidal features produced by minor
mergers in the redshift range $1<z<2$. Figures
\ref{fig:relaxed_SG_examples} and \ref{fig:disturbed_SG_examples}
show images of typical R-SGs and D-SGs respectively. Note that our
D-SGs do not include `close pair' systems, where there are two
well-separated galaxies that have not yet coalesced.


\subsection{Stellar masses and rest-frame photometry} Stellar
masses and rest-frame photometry are calculated via spectral
energy distribution (SED) fitting. The WFC3/ACS photometry of each
individual galaxy is compared to a large library of synthetic
photometry, constructed using exponentially-decaying star
formation histories (SFHs), each described by a stellar mass
($M_*$), age ($T$), e-folding timescale ($\tau$), metallicity
($Z$) and internal extinction ($E_{B-V}$). We vary $T$ between
0.05 Gyrs and the look-back time to $z=20$ in the rest-frame of
the galaxy, $\tau$ between 0.01 Gyrs (approximately an
instantaneous burst) and 9 Gyrs (approximately constant star
formation), $Z$ between 0.1 Z$_{\odot}$ and 2.5 Z$_{\odot}$ and
$E_{B-V}$ between 0 and 1 mag. Synthetic magnitudes are generated
by folding the model SFHs with the stellar models of
\citet{BC2003} through the correct WFC3 and ACS filter
throughputs(assuming a Chabrier initial mass function), with dust
attenuation applied following \citet{Calzetti2000}. The likelihood
of each model, $\exp (-\chi^2/2)$, is calculated using the value
of $\chi^2$, computed in the standard way. Estimates for the free
parameters such as stellar mass are derived by marginalising each
parameter from the joint probability distribution, to extract its
one-dimensional probability density function (PDF). We use the
median of this PDF as the best estimate of the parameter in
question, with the 25 and 75 percentile values (which enclose 50\%
of the probability) yielding an associated uncertainty. The
derived stellar masses are uncertain by $\sim$0.2 dex. The
K-corrections required to construct rest-frame photometry for each
galaxy are calculated using the best-fit model SED (i.e. where the
value of $\chi^2$ is a minimum).


\subsection{Sizes}
Galaxy effective radii ($R_e$) are calculated via the WFC3 F160W
images, using \texttt{GALAPAGOS} \citep{Barden2012}, an IDL-based
pipeline for running \texttt{SEXTRACTOR} \citep{Bertin1996} and
\texttt{GALFIT} \citep{Peng2002} together. Individual galaxies are
fitted with a 2D Sersic profile, using the default
\texttt{GALAPAGOS} parameters \citep[see][]{Haussler2007}. We use
circularised radii $R_e = r^{\texttt{FIT}}_e \times \sqrt{b/a}$ as
our size estimator.

\begin{figure}
\begin{center}
\includegraphics[width=3in]{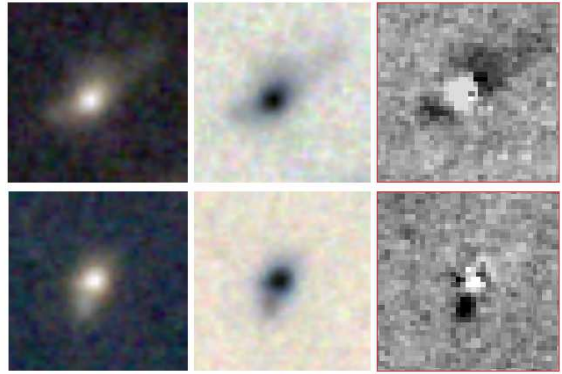}
\caption{Examples of disturbed SGs in our dataset. We show the
$YJH$ colour composite image (left column) and its negative
(middle column) and the residual image from the \texttt{GALFIT}
fits (right column). Numerical simulations of mergers in the
redshift of our study indicate that only tidal features from major
mergers (mass ratios $>$ 1:4) will be visible in the ERS image
(features from minor mergers being too faint). These disturbed SGs
are therefore like to be major-merger remnants.}
\label{fig:disturbed_SG_examples}
\end{center}
\end{figure}

\begin{figure*}
\begin{minipage}{172mm}
\begin{center}
$\begin{array}{cc}
\includegraphics[width=1.5in]{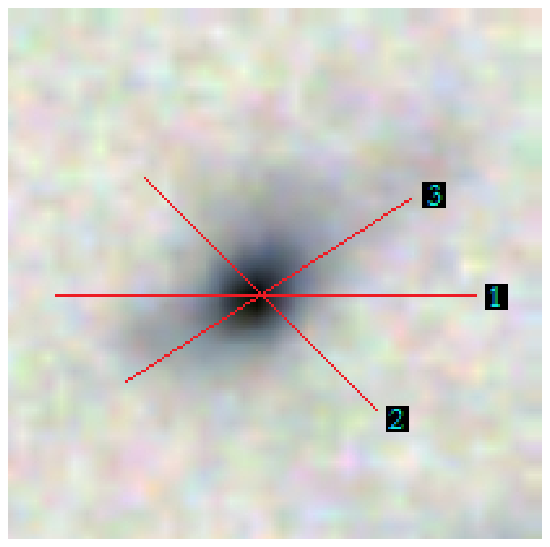} & \includegraphics[width=2.5in]{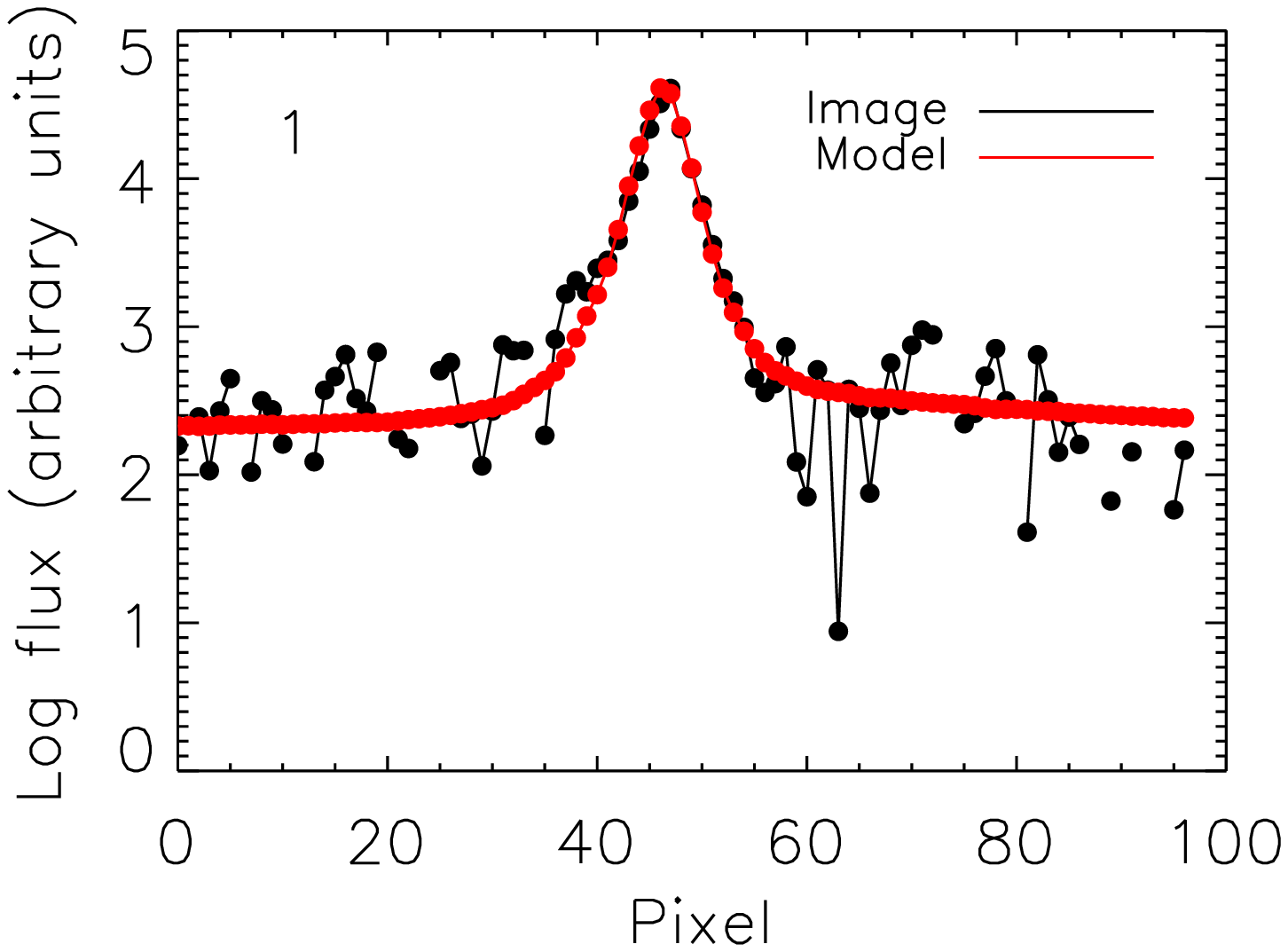}\\
\includegraphics[width=2.5in]{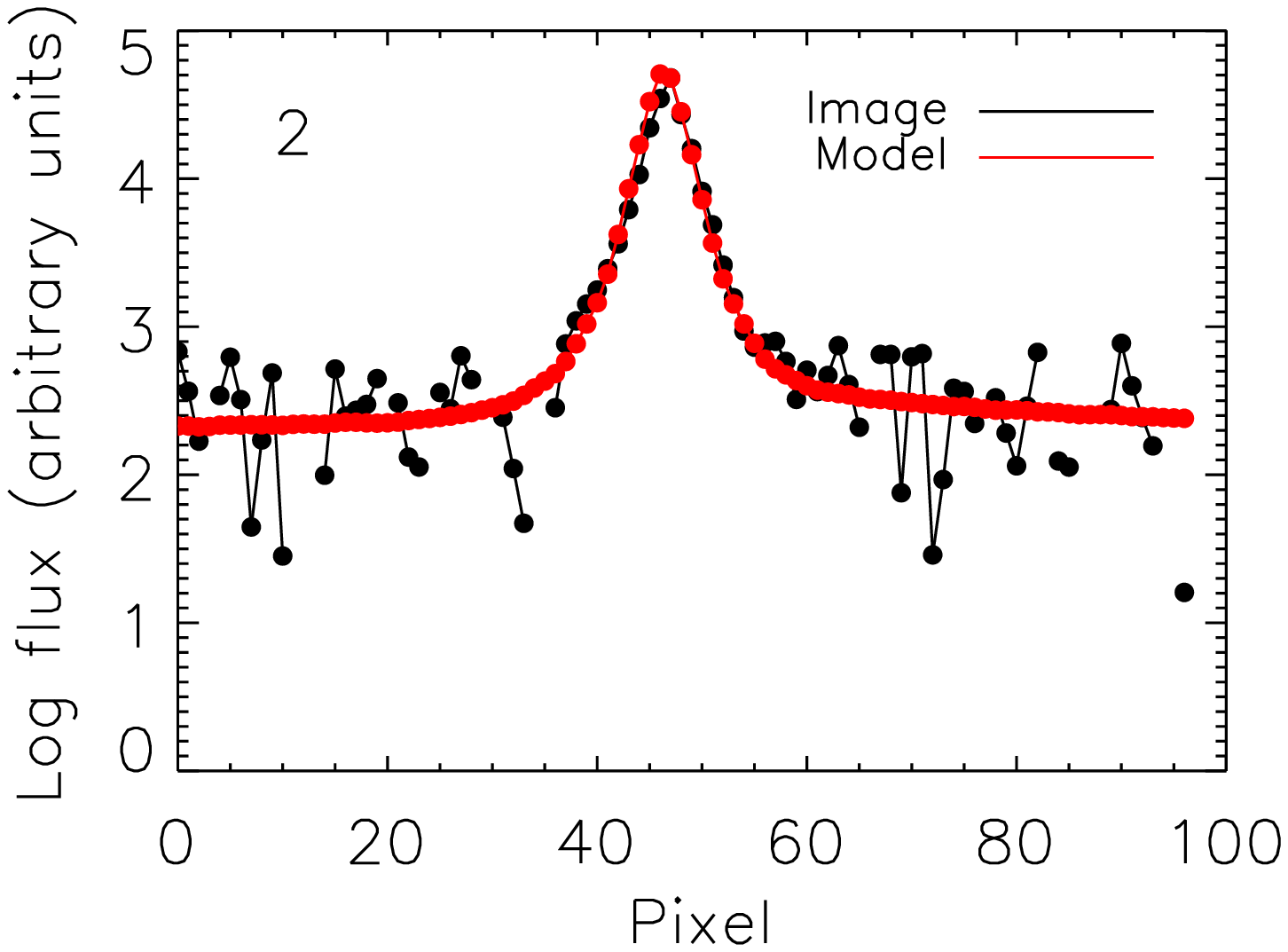} &
\includegraphics[width=2.5in]{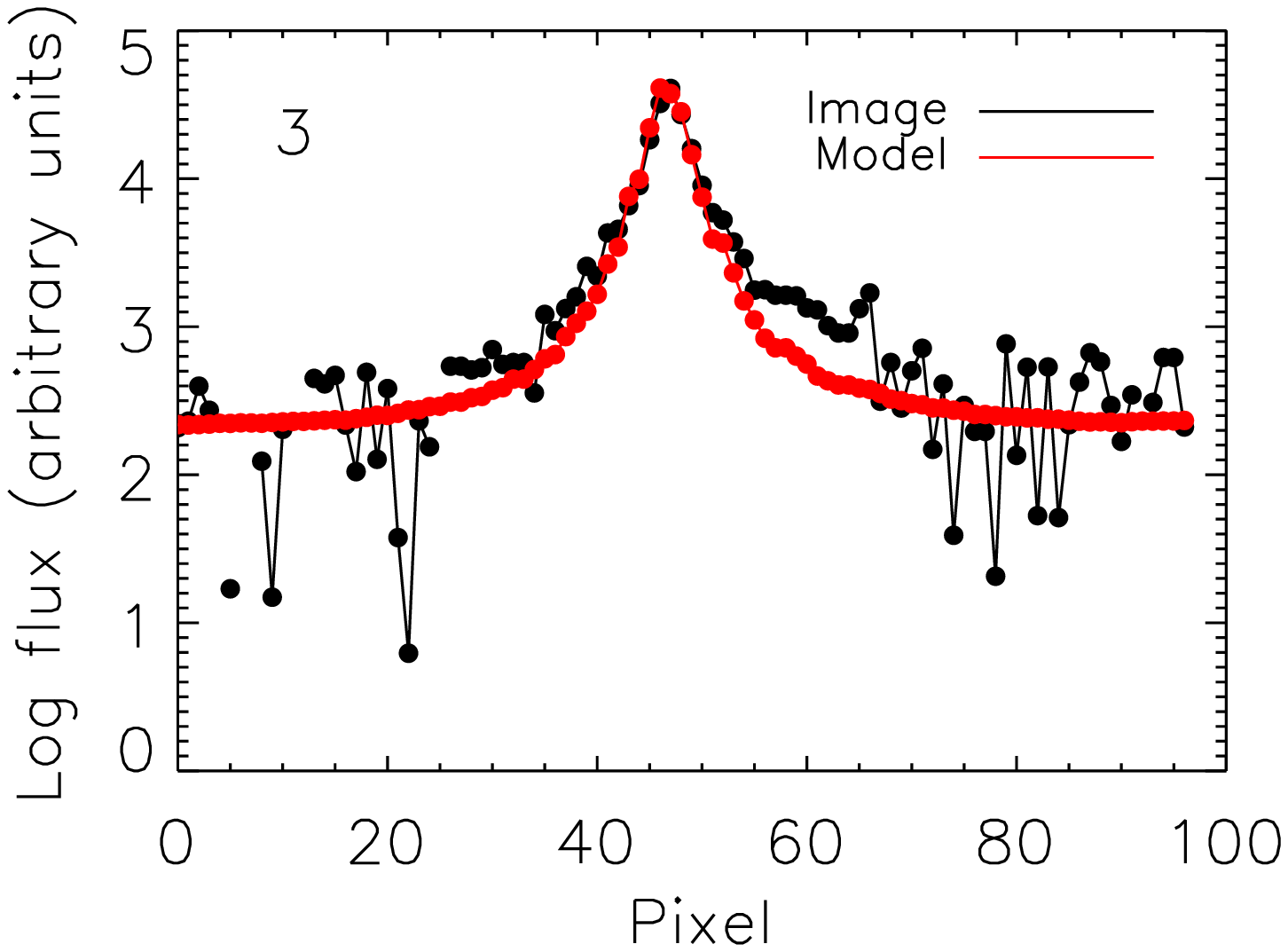}
\end{array}$
\caption{Flux profiles of a disturbed SG (top left; see Figure 2 for
original image). Note that the fluxes are on a log scale. We show profiles
taken perpendicular to the tidal feature (top right and bottom left) and
along the feature (bottom right). Black indicates the original $H$-band
galaxy image and red indicates the best-fit \texttt{GALFIT} model. The
number of the profile axis (see red lines in the top-left panel) is
indicated in each plot.}
\label{fig:slices}
\end{center}
\end{minipage}
\end{figure*}

It is important to ensure that the faint tidal features around our
D-SGs are not affecting their fitted sizes. While the features are
brighter than the background (especially in the co-added images)
and thus detectable by eye, they host a negligible fraction (a few
percent) of the total luminosity of the system. As the residuals
in Figure \ref{fig:disturbed_SG_examples} show, the tidal features
do not get included in the model fits and thus do not affect the
derived effective radii. This is reflected in the lack of a trend
between $R_e$ and the physical extent of the features, again
indicating that they do not affect the measured sizes. In Figure
\ref{fig:slices}, we demonstrate these points more explicitly, by
presenting the flux profiles of a D-SG that has strong tidal
features extending to several effective radii. If the effective
radius was being overestimated due to the tidal features, then the
fitted model \emph{profiles} would be broader than that of the
original galaxy. However, regardless of whether profiles are
plotted perpendicular to (top right \& bottom left) or along
(bottom right) the tidal features, it is clear that the model
profiles are not preferentially broader, consistent with the fact
that the faint features are not included in the model fits. This
behaviour is the same for all D-SGs in our sample and is
essentially due to the fact that the total luminosity in the
features is a negligible fraction of that in the whole system.
Note that the fluxes in Figure \ref{fig:slices} are shown on a log
scale.

It is worth noting here that past size-evolution studies typically
do not consider the effect of tidal features, which become
increasingly common at higher redshift, especially at $z>1$
\citep[e.g.][]{Kaviraj2013a}. Nevertheless, our analysis indicates
that the derived $R_e$ values in such studies are also likely to
be reliable.

\section{The role of major mergers in driving size growth in spheroids}
\begin{figure}
\begin{center}
\includegraphics[width=3in]{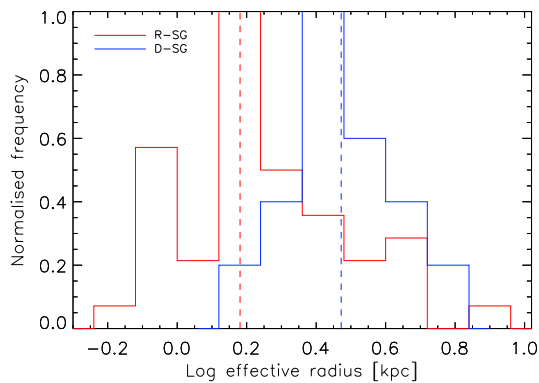}\\
\caption{The size distribution of the D-SG and R-SG populations.
The R-SGs are restricted to the D-SG mass range
($M_*$$<10^{10.7}$M$_{\odot}$).} \label{fig:size_dist}
\end{center}
\end{figure}

\begin{figure}
\includegraphics[width=3.3in]{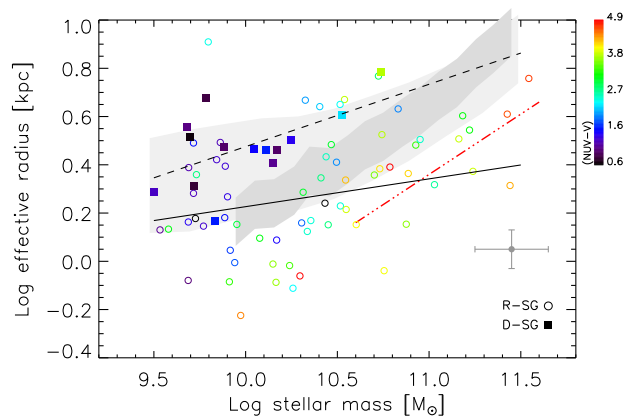}
\caption{The $R_e$-$M_*$ relation of SGs at $1<z<2$. Circles
indicate relaxed SGs (R-SGs) and squares indicate disturbed SGs
(D-SGs). The dashed and solid lines show linear best-fits to the
D-SGs and R-SGs (restricted to the D-SG mass range,
$M_*$$<10^{10.7}$M$_{\odot}$) respectively. The light and dark
grey shaded regions indicates the the location of local SGs taken
from \citet{Shen2003} and \citet{Bernardi2012} respectively. The
red dashed dotted line indicates the best-fit line from Newman et
al. (2012).} \label{fig:re_props}
\end{figure}

In Figure \ref{fig:size_dist} we present the size distribution of
the D-SG and R-SG populations. The R-SGs are restricted to the
D-SG mass range ($M_*$$<10^{10.7}$M$_{\odot}$). Not unexpectedly,
D-SGs are larger than their relaxed counterparts. The difference
in the median values of the two distributions is a factor of 2.
Since our SGs span a range in stellar mass, we plot, in Figure
\ref{fig:re_props}, the effective radius as a function of stellar
mass. Circles represent R-SGs and squares represent D-SGs. The
colour coding indicates the star-formation-sensitive rest-frame
$(NUV-V)$ colour. The $NUV$ filter, taken from the GALEX
filterset, is centred at 2300 \AA. The light and dark grey shaded
areas indicate the region occupied by local SGs, taken from
\citet{Shen2003} and \citet{Bernardi2012} respectively.
{\color{black}The measured sizes of our SGs are similar to those
in previous studies that have probed similar masses and epochs
\citep[see
e.g.][]{Damjanov2011,Cimatti2012,Newman2012,Cassata2013}. We
indicate the best-fit line from Newman et al. (2012) using a red
dashed dotted line and note that the low-mass end of the size-mass
relation (which is the focus of our study) appears to be
consistent with the high-mass end that has been studied in the
recent literature.}

Notwithstanding the scatter in the $R_e$-$M_*$ relation, the D-SGs
cluster towards the upper envelope of the relation, largely
independent of stellar mass. To estimate major-merger-driven size
growth, we first calculate linear best fits to the D-SGs and R-SGs
(dashed and solid lines respectively), restricted to the D-SG mass
range ($M_*$$<10^{10.7}$M$_{\odot}$). The difference between these
linear fits then offers a mean statistical estimate of the size
growth plausibly induced by the major-merger process. We find a
size growth of around a factor of 2, with a weak trend with
stellar mass.

{\color{black}The D-SG population is consistent with the upper
envelope of the local size-mass relation, indicating that in the
mass range considered here, major mergers may play a significant
role in bringing at least some SGs onto the local size-mass
relation. Combining the observed (factor x2) size growth with the
expected major-merger history of SGs enables us to further explore
the size evolution induced by this process on the SG population as
a whole. Empirical estimates of the major-merger history indicate,
on average, $\sim$0.3 major mergers after $z\sim1$
\citep[e.g.][]{Lopez2009,Lotz2011}. Given the redshift of the
systems studied here, one expects around a third of the SGs to
experience around one more major merger as they evolve to the
present day. Of the 60 SGs in this mass range, 13 D-SGs and 20
R-SGs are already consistent with the local relation. 22 R-SGs are
below the local relation - if 30\% of these undergo a major merger
at $z<1$ and experience a factor $\times$2 size increase, then a
further 7 SGs will end up being consistent with the local
relation. This implies that, for a reasonably high fraction (40/60
= 66\%) of SGs in this mass range, major mergers could have played
a significant role in their cosmic size evolution.


This process will also move the descendants of some our D-SGs
\emph{into} the mass selection typically employed in
size-evolution studies ($M_*>10^{10.5}$ M$_{\odot}$). As noted in
the introduction, recent work on the evolution of the SG mass-size
relation indicates that, to produce the observed size growth in
the $M_*>10^{10.5}$ M$_{\odot}$ regime, one requires new,
systematically larger systems to directly enter this mass
selection at $z<1$ \citep[`progenitor bias', see
e.g.][]{Newman2012,Carollo2013,Patel2013}. Some of our D-SGs are
natural candidates for being the progenitors of these new systems,
implying that the size evolution of at least some local SGs with
$M_*>10^{10.5}$ M$_{\odot}$ could also be heavily influenced by
major mergers alone.

It is worth noting here that the D-SGs are bluer than their
relaxed counterparts in $(NUV-V)$, indicating that the major
mergers that produced their tidal features were not completely
dry. From an empirical standpoint this is expected, since
\emph{all} galaxies at these redshifts host some star formation.
For example, \citet{Kaviraj2013a} have used rest-frame UV-optical
colours to show that none of the massive galaxies at these epochs,
not even SGs, are consistent with passively-evolving populations.
Thus, while theoretical work often invokes `dry' (gas-free)
merging, typically based on spheroid-spheroid interactions, to
explain size growth, observational studies indicate that merger
events (major or minor) at these epochs are not dry. Nevertheless,
the colours of our D-SGs indicate that major mergers that are not
completely gas-free are also capable of increasing SG sizes by
factors of $\sim$2.

In this context it is also worth noting that major mergers
involving high gas fractions (e.g. $>$30\%) may create remnants
that have substantial disk components \citep[e.g.][]{Hopkins2009}.
It is, therefore, instructive to explore whether our D-SGs will
indeed eventually join the R-SG population, or whether some of
these systems may end up as disky remnants. While a robust answer
requires a direct measurement of the cold gas mass, from e.g.
ALMA, some qualitative insight can be gained by assuming that the
D-SGs follow the global Schmidt-Kennicutt law:

\begin{equation}
\psi = (\epsilon/\tau_{dyn}).M_g,
\end{equation}

where $\psi$ is the star formation rate, $\epsilon$ is the star
formation efficiency, $\tau_{dyn}$ is the dynamical timescale of
the system and $M_g$ is the mass of the {\color{black}cold} gas
reservoir.

Using the SED-fitted SFRs and stellar masses for our D-SGs, we
`invert' the Schmidt-Kennicutt law, assuming $\tau_{dyn}$ = 0.1
Gyrs (which seems reasonable for systems at this redshift, see
e.g. Hopkins et al. 2009), which yields a value for $M_g$.
Dividing $M_g$ by the stellar mass then yields an estimate of the
gas fraction for each D-SG. The values we derive are all less than
10\%. In this regime disk regrowth is unlikely, implying that the
D-SGs studied here are likely to join the R-SG population when the
tidal features due to the recent merger fade. It is worth noting
that the factor of 2 size increase is then consistent with the
expectation of theoretical work that also assumes gas-poor
progenitors \citep[e.g.][]{Ciotti2007,Naab2009,Hilz2012}.

Finally, although our D-SGs are, by construction, `post-merger'
systems, it is useful to study the morphological properties of
their progenitor systems, which could either be a major merger
between two spirals or one between a spiral and an SG. Late-type
galaxies are typically several factors larger than their
early-type counterparts \citep[see e.g.][]{vanderwel2014}. While
mergers between late-type galaxies are expected to produce
remnants that are somewhat smaller than their progenitors, the
reduction in size (10-20\%, e.g. Cox et al. 2006) does not appear
sufficient to produce remnants that are close to the early-type
population. The estimated gas fractions of less than 10\% in our
D-SGs also indicate that they are inconsistent with gas-rich major
mergers between late-type galaxies, that are likely to leave
remnants that are more gas-rich \citep[e.g.][]{Hopkins2009}. It is
likely, therefore, that the progenitor systems of our D-SGs were
typically major mergers between an SG and a spiral (late-type)
system of similar mass.}

\section{Summary}
We have studied the structural properties of
{\color{black}intermediate-mass ($10^{9.5}$
M$_{\odot}$$<M_*<10^{10.7}$ M$_{\odot}$)}, high redshift ($1<z<2$)
spheroidal galaxies (SGs) in the WFC3 Early-Release Science
programme, to derive an empirical estimate for the size growth
induced by the major-merger process in this mass range. We have
visually split our SG sample into galaxies that are relaxed and
those that exhibit tidal features indicative of a recent merger.
Numerical simulations indicate that, given the depth of the ERS
images, only tidal features due to \emph{major} mergers (mass
ratios $<1:4$) are likely to be visible in the redshift range
$1<z<2$ (features due to more minor mergers being too faint for
detection). In other words, the tidally `disturbed' SGs (D-SGs)
that exhibit such features have experienced a recent major merger,
making them a valuable set of objects with which to
\emph{empirically} estimate the structural impact of the
major-merger process.

We find that the D-SGs are offset in size from their relaxed
counterparts, lying towards the upper envelope of the radius
($R_e$) -- stellar mass ($M_*$) relation at these epochs,
indicating that in the mass range considered here, major mergers
may play a significant role in bringing at least some SGs onto the
local size-mass relation. The median ratio of the effective radii
of the D-SGs to that of their relaxed counterparts is $\sim$2,
with a weak trend with galaxy stellar mass. It is worth noting
that, while models typically invoke \emph{dry} mergers to explain
size growth, empirical work, such as the one presented here,
indicates that major-merger remnants at these epochs are not
completely dry. {\color{black} While our estimates of the gas
fraction in these systems indicates values less than around 10\%,
our results indicates that such wet mergers at these redshifts are
capable of inducing size growth of around a factor of 2, at least
in the mass range studied here. The low gas fractions, combined
with the comparative position of late-type galaxies on the
size-mass relation, suggests that our D-SGs are remnants of major
mergers between SGs and spirals, and will transform into relaxed
SGs when their tidal features fade.}

{\color{black}We have combined the observed (factor x2) size
growth with the expected major-merger history of SGs to study how
the major merger process may affect SG size evolution in our mass
range of interest. Estimates of the major-merger history indicate,
on average, $\sim$0.3 major mergers after $z\sim1$. Given the
redshift of the systems studied here, we have estimated that at
$z\sim0$, around two-thirds of SGs within our mass range could
have had their size evolution driven primarily by major mergers
alone.}


\section*{Acknowledgements}
We are grateful to the anonymous referee for a useful report which
improved the original manuscript. We thank Ignacio Trujillo,
Emanuele Daddi, Chris Conselice and Richard Ellis for interesting
discussions. S.K. acknowledges a Senior Research Fellowship from
Worcester College Oxford. This paper is based on Early Release
Science observations made by the WFC3 Scientific Oversight
Committee. We are grateful to the Director of the STScI for
awarding Director's Discretionary time and deeply indebted to the
brave astronauts of STS-125 for rejuvenating HST. Support for HST
program 11359 was provided by NASA, through grant GO-11359 from
the STScI, which is operated by the Association of Universities
for Research in Astronomy, Inc., under NASA contract NAS 5-26555.
R.A.W. acknowledges support from NASA JWST Interdisciplinary
Scientist grant NAG5-12460 from GSFC.

\small \nocite{Windhorst2011,Trujillo2011,Cox2006}
\bibliographystyle{mn2e}
\bibliography{references}


\end{document}